# Fiber laser based stimulated Raman photothermal microscopy with long working distance optics

Xiaowei Ge[a,†], Yifan Zhu[a,†], Dingcheng Sun[b], Hongli Ni[a], Yueming Li[a], Chinmayee V. Prabhu Dessai[b], Ji-Xin Cheng [a,b,c,d]*

[a]Department of Electrical & Computer Engineering, Boston University, Boston, Massachusetts, USA.

[b]Department of Biomedical Engineering, Boston University, Boston, Massachusetts, USA.

[c]Department of Chemistry, Boston University, Boston, Massachusetts, USA.

[d]Photonics Center, Boston University, Boston, Massachusetts, USA.

[†]These authors have contributed equally to this work

**Corresponding authors**: jxcheng@bu.edu

# Abstract

Stimulated Raman scattering (SRS) microscopy is a highly sensitive chemical imaging technique. However, the broader application of SRS has been limited by two key challenges: the reliance on low-noise but bulky solid-state laser sources and stringent sample requirements necessitated by high numerical aperture (NA) optics. Here, we present a fiber laser based stimulated Raman photothermal (SRP) microscope that addresses these limitations. While appreciating the portability and compactness of a noisy source, fiber laser SRP enables a two-order-of-magnitude improvement in signal to noise ratio over fiber laser SRS without balance detection. Furthermore, with the use of low NA, long working distance optics for signal collection, SRP expands the allowed sample space from millimeters to centimeters, which diversifies the sample formats to multi-well plates and thick tissues. The sensitivity and imaging depth are further amplified by using urea for both thermal enhancement and tissue clearance. Together, fiber laser SRP microscopy provides a robust, user-friendly platform for diverse applications.

# 1 Introduction

Bond-selective chemical imaging technologies are opening a new window to scrutinize molecular events in biological, environmental, and energy related systems. Of the various modalities, stimulated Raman scattering (SRS) microscopy allows high-throughput vibrational imaging, offering linearity of signal intensity with chemical bond concentration [1,2]. SRS microscope uses two synchronized pump and Stokes laser beams with their energy difference matching the Raman-active bond vibrational frequency [3]. Life science applications of SRS imaging include rapid label-free tissue diagnostics via stimulated Raman histology [3–5], imaging of altered cell metabolism in cancers, aging and neurodegenerative diseases [6–10], resolving microbial heterogeneity and drug response in a microbiome [11–14], and super-multiplex bio-imaging [15,16].

Despite these advances, broader use of SRS microscopy is limited by the laser noise and the strict requirement for high numerical aperture (NA) light collection [17]. The shot noise originating from the local oscillator in SRS often dominates when a solid-state laser is used as the excitation source. With the recent trend of translational applications, ultrafast fiber lasers offer a solution with its portability and insensitivity to the environment [18]. However, the laser intensity noise in fiber lasers poses a more prominent problem compared with shot noise, which significantly limits the imaging sensitivity [19]. To address the noise problem, innovations such as balanced detection and quantum enhanced SRS have been developed [19–24]. The balance detection approach could suppress the laser noise with the trade of increasing the shot noise by 3 dB [19]. However, this approach involves a complicated setup and is sensitive to environmental electronic noise, often yielding suboptimal performance, which limited the application to dense tissue samples. Quantum enhanced SRS with squeezed photons is a nice way to suppress the shot noise and 3.6 dB improvement was achieved [22]. However, with the setup complexity and vulnerability to optical loss characteristics, it is not capable of wide implementation, especially for translational usage in clinic.

The second challenge in SRS microscopy lies in the requirement for a high NA objective for excitation and a high NA condenser for light collection. The high NA objective is essential for achieving sufficient SRS intensity and subcellular resolution, while the high NA condenser ensures the full capture of the SRS signal, minimizes the cross-phase modulation (XPM), and maintaining the spectral fidelity [25,26]. Typically, the NAs greater than 1 are used, necessitating the use of immersion objectives and condensers with media such as water or oil. This setup restricts sample thickness to the millimeter scale and eliminates any air gap, limiting the adaptability of SRS microscopy to diverse sample types [27], including flow cytometry, multi-well plates, and cell ejection chips [27–29]. Furthermore, the intricate alignment required to overlap the tight foci of the objective and condenser during sample replacement is time-consuming and demands significant expertise.

Recently developed stimulated Raman photothermal (SRP) microscope [30] is a new form of stimulated Raman microscopy that leverages the high sensitivity of photothermal detection [30,31]. In the SRS process, the pump and Stokes beams undergo intensity loss and gain, respectively, with the energy difference ($\hbar\omega_p - \hbar\omega_S$) being transferred to the target molecule. The subsequent vibrational relaxation generates heat, which negatively modulates the refractive index in the focus. In an SRP microscope, a third beam is employed to measure the resulting thermal lensing effect. Versatile applications on viral particles, cells, and tissues were demonstrated [30]. However, current SRP microscope deploys a bulky free-space optical parametric oscillator (OPO) laser not suitable for clinical use. In addition, a high NA objective and an oil condenser was used for signal collection, making the SRP system not applicable to imaging live cells cultured in multi-well plates for high throughput measurements.

In this work, we report a fiber laser SRP microscope that offers both high sensitivity with operational simplicity, leveraging the three key advantages of SRP microscopy. First, the noise in SRP measurements is independent of the pump or Stokes beam, making the system highly resilient to the inherent laser noise of the compact fiber laser sources. This eliminates the need for complex noise cancellation setups while enabling highly sensitive detection with fiber laser system. Second, SRP detection does not require high NA light collection. Accordingly, our fiber laser SRP system employs an inverted microscope configuration with low NA, long working distance light collection optics, enabling contact-free measurement of versatile samples and greatly improving the operational efficiency. Third, SRP enhances vibrational signal contrast and significantly boosts signal levels through thermal lensing detection, particularly in thermal enhancement media [30,32]. We found that urea, a tissue clearing agent, acts as an effective thermal enhancement medium for SRP, providing both signal amplification and imaging depth improvement [33,34]. Collectively, these advances establish a new platform technology with significant potential for both basic and clinical research.

## 2 Results

## 2.1 A fiber laser SRP microscope

We developed the design (**Fig. 1a**) to utilize the advantages of SRP technology to push this next-generation chemical imaging microscope toward clinical applications. For the laser source, we selected a compact, dual output, picosecond tunable fiber laser (Picus Duo, Refined Laser Systems) [35]. The fiber laser system generates synchronized and tunable pump and Stokes beams, fast sweeping molecular vibrational frequency ranging from 700 cm$^{-1}$ to 3100 cm$^{-1}$ without any beam pointing issues (detailed benefits in Materials and methods). After the output of the laser, both beams were expanded with a pair of lenses composite Keplerian beam expander. At the beam expander focus, acousto-optic modulators (AOM) modulated both beam with aligned frequency and phase, controlled by a function generator. The modulation on both beams can be controlled independently on and off, to flexibly adapt to avoid pump or Stokes non-resonant absorption-induced photothermal background contribution during SRP measurement. Then the two beams were colinearly combined together with a dichroic mirror. A 765 nm continuous wave (CW) probe laser, offering significantly improved noise performance compared to the fiber laser source, was first expanded through a Keplerian beam expander. The second lens of this expander was mounted on a translational stage, allowing control over the distance between the two lenses and, consequently, the divergence of the probe laser and its relative focal position to the pump and Stokes beams foci of the objective. The offset allows the probe beam to experience a spatially varying phase shift, which maximizes the interaction between the thermal induced refractive index gradient and the probe beam, thus amplifying the thermal lensing effect and improving the detection of small temperature variations or weak absorptions [30,31,36]. The probe laser was then combined with the pump and Stokes beams using a polarization beam splitter and the same dichroic mirror used for pump and Stokes beams. The three collimated beams were directed into a scanning unit equipped with a pair of galvo mirrors, conjugated to the back aperture of the objective in the microscope by a four focal length relay (4f) system, to facilitate laser scanning imaging.

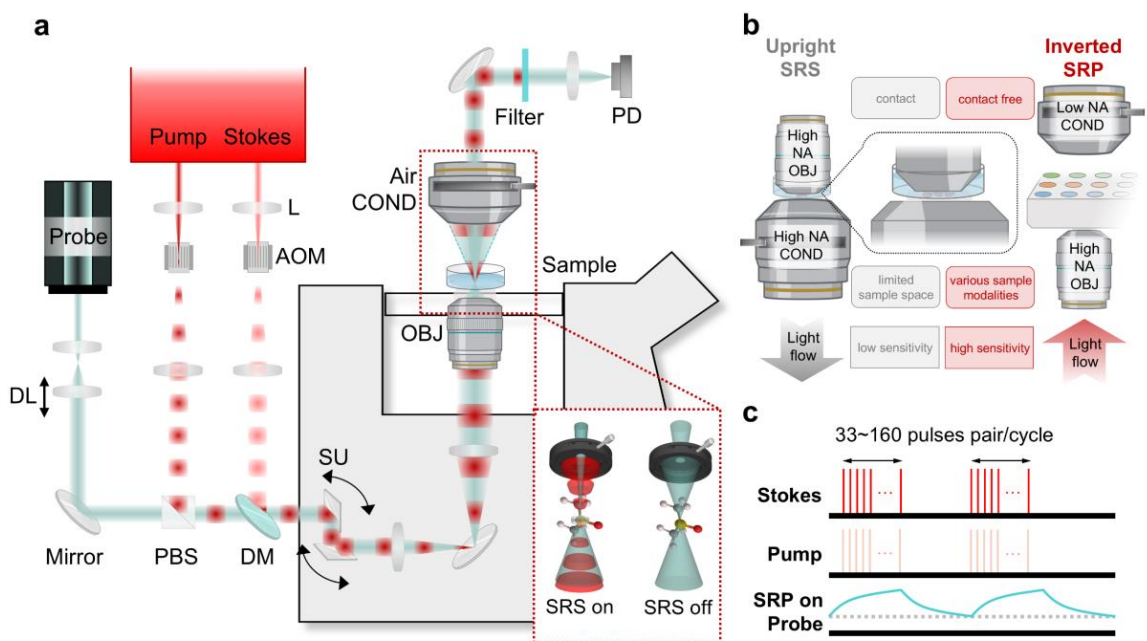

**Fig. 1** Schematic of a fiber laser SRP microscope. **a**. Fiber laser SRP setup. L: lens. AOM: acousto-optic modulator. TL: translational lens. BS: polarized beam splitter. DM: dichroic mirror. SU: scanning unit. OBJ: objective. COND: condenser. PD: photodiode. The red dashed line box illustrated the SRP optical contrast by detailed light path alternation due to the presence of the SRP induced thermal lens. **b**. Comparison with upright SRS and inverted implemented SRP. **c**. Modulated Stokes and pump pulses induced SRP signal detected by the probe beam. The SRP signal shows a clear thermal heat accumulation and decay signature.

To configure the microscope favorably for clinical use, versatile sample types, and ensure the compatibility with commercialized microscopy systems such as confocal fluorescence microscopes, an inverted microscope configuration was implemented [37,38]. To leverage the advantage that SRP does not require high NA collection, a low NA, long working distance air condenser was used on the top for light collection. The long working distance simplifies sample change without necessitating optical adjustments. It also accommodates thick samples, dry samples or more advanced multi-well cell culture plates (**Fig. S1**), ensuring that the SRP microscope is well suited for both research and clinical applications. This configuration is advantageous over SRS microscopes (**Fig. 1b**), which typically requires high NA oil condenser for signal collection, to avoid cross-phase modulation [25]. In the inverted SRS microscope configuration, oil condenser is on the top. It impedes the measurement on cell culture dishes containing aqueous solutions which are crucial for biology study. While in the upright SRS microscope configuration, the system will be limited to water objective when measuring cell culture dish samples (**Fig. 1b**) and tend to cause sample contaminations. Under both schemes, SRS needs to readjust the optics each time changing the sample, due to the narrow sample space caused by the two high NA optics. More details are provided in a later session. The microscope includes a 3-dimension motorized sample stage, enabling automated sample imaging and large-area mapping. After the light collection by the low NA condenser, the probe beam was filtered out optically and electronically, and finally input into a lock-in amplifier to detect the modulation transfer from the pump and/or Stokes beams (**Fig. 1c**).

## 2.2 Thermal lensing simulation to determine the optimal NA for SRP signal collection

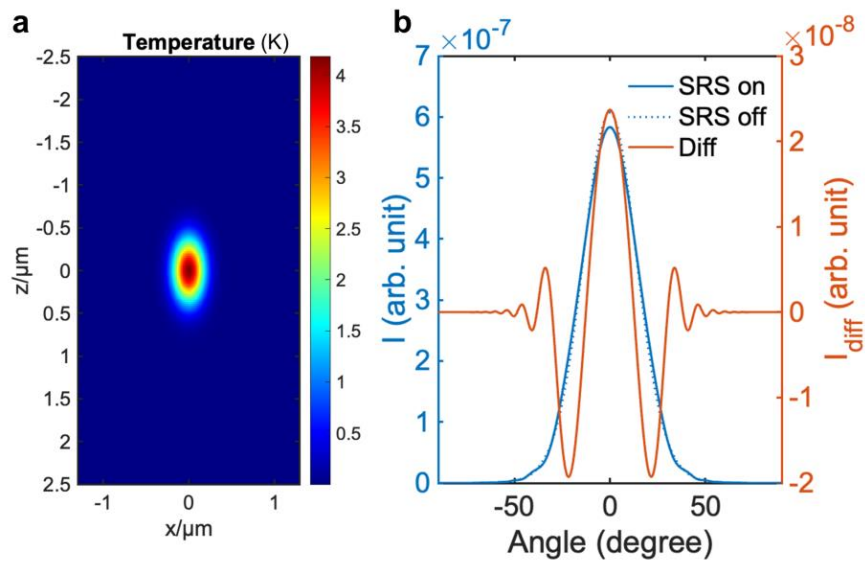

**Fig. 2** Simulated SRP induced temperature increasement and probe light far field distribution. **a**. SRS induced focus temperature increase with 28 mW pump and 180 mW stokes on sample for 8 μs. **b**. Far field distribution of the probe beam with SRS on (blue solid line), off (blue dashed line), and difference (Diff) between on and off states (red line).

To examine the NA required for optimal contrast in SRP microscopy and to estimate the performance of SRP under the conditions provided by a fiber laser system, we conducted thermal lensing simulations (**Fig. 2**). To provide the energy deposition information for SRP simulation, first, we measured the SRS modulation depth in a fiber laser system at its normal operation power, which usually does not cause any biological sample burning or detector saturation. With the fiber laser 40 MHz dual output, an unmodulated 28 mW pump beam and a 90 mW 50%-duty cycle modulated Stokes beam on the sample, the SRS modulation on pump beam was measured to be 0.72% when targeting the carbon-hydrogen (C-H) bond in pure dimethyl sulfoxide (DMSO) at 2912 cm$^{-1}$ (**Fig. S2a**). This corresponds to an energy deposition of 2.33 pJ per pulse pair with this 40 MHz fiber laser source.

Using this energy deposition through stimulated Raman vibrational absorption, we calculated the induced temperature increase based on the thermal properties of DMSO (**Table S1**). The results indicate that the temperature increase can reach up to 4 Kelvin at the center of the focus (**Fig. 2a**), demonstrating significant thermal effects induced by the SRS process using a fiber laser system. By converting the temperature change to a refractive index change, we were able to calculate changes in the far-field probe light distribution resulting from thermal lens effects induced by SRS [30]. The far-field probe light distribution was then simulated under conditions of SRS on and SRS off to determine the necessary NA for effective SRP measurement (**Fig. 2b**, see details in Materials and methods section). The simulations revealed that an NA of 0.32 is required to achieve the maximum contrast between the SRS on and off conditions in the DMSO C-H measurement. With this configuration, SRP measurement on the probed beam could have a modulation depth of 3.3% (**Fig. S2b**) at 125 kHz modulation frequency, which shows a decent modulation depth increasement by implementing SRP on the fiber laser system.

This information is critical for guiding the configuration of the SRP system to ensure optimal signal detection and contrast, thereby enhancing the overall performance of SRP microscopy for various applications. The condenser used in our setup is a 28 mm long working distance air condenser, with a tunable NA up to 0.55 (Nikon), providing flexibility for testing various samples while maintaining the long working distance feature.

### 2.3 Performance of the fiber laser SRP microscope

To optimize the measurement conditions for our SRP system and compare its sensitivity with previous fiber laser-based, autobalanced detected SRS [21], we conducted a series of tests to characterize SRP sensitivity. Specifically, we examined the SRP on DMSO for signal-to-noise ratio (SNR) dependency on duty cycle and modulation frequency. With conserved laser peak power, the optimal performance was achieved with a 50% duty cycle. Notably, the laser peak power is conserved in this comparison, due to the relatively low available laser power and negligible photodamage of the fiber laser. In this case, higher excitation duty cycle, although leads to insufficient cooling, could provide higher total energy deposition hence better signal. The optimal modulation frequency in fiber laser SRP is 600 kHz, influenced by the probe laser's performance. These conditions provided the highest SNR, demonstrating the system's sensitivity and potential for various applications (**Fig. 3a,b**). The comparison with fiber laser SRS showed that our SRP system offers significantly improved sensitivity, achieving an ~12-fold improvement over autobalanced detected SRS and ~105-fold improvement over SRS in terms of SNR for DMSO (**Fig. S3**). With the gradient dilution measurement, we determined the C-H and carbon-deuterium bond (C-D) region limit of detection (LOD) by DMSO and its deuterated counterpart (DMSO-d6) when diluting with each other (**Fig. 3c,d**). The LOD of a 20 μs/wavenumber measurement in C-H region is 11.13 mM with DMSO diluted with DMSO-d6. The LOD in C-D region is 13.77 mM with DMSO-d6 diluted with DMSO.

SRP microscopy benefits from the confined nonlinear excitation at the SRS focus and the third beam probing geometry, leading to improved resolution that can reveal fine structures in biological samples. To characterize the system's resolution, we measured and deconvoluted the point spread function using 100 nm polymethyl methacrylate (PMMA) beads at 2950 cm$^{-1}$ in the C-H region. The results demonstrated a spatial resolution of 240 nm and an axial resolution of 0.82 μm (**Fig. 3e-h**, **Fig. S4**). These results are consistent with the high-resolution capabilities observed in the previous study [30], where SRP showed superior spatial resolution compared to SRS, providing sharper imaging of fine structures such as tiny microbiome and subcellular organelles. This high resolution underscores the capability of SRP microscopy to provide detailed imaging of small features, making it a valuable tool for studying intricate biological structures and potentially enhancing the understanding of cellular and subcellular processes.

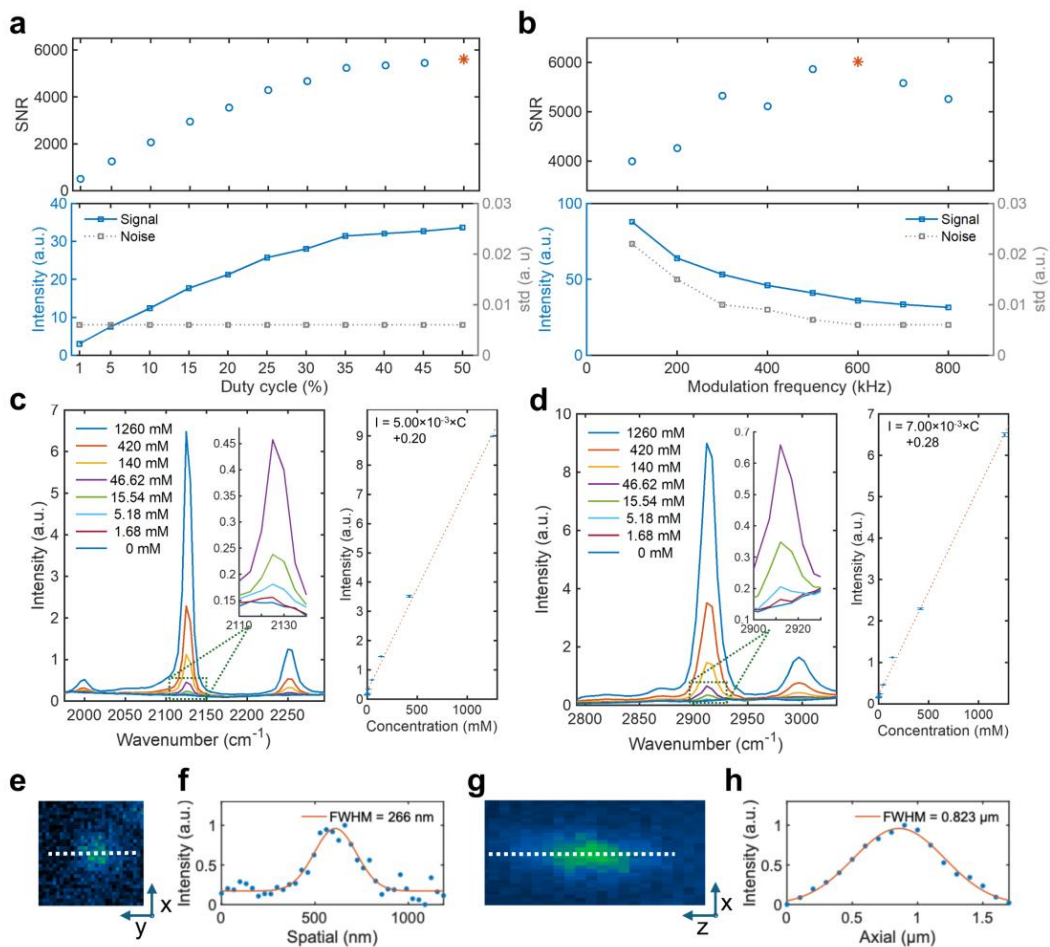

**Fig. 3** Sensitivity and spatial resolution of fiber laser SRP. **a**. SRP signal duty cycle and modulation frequency dependency. **b**. SNR under different modulation frequency. **c**. Spectral fidelity (left) and intensity-concentration relation (right) of serial diluted dimethyl sulphoxide-d6 (DMSO-d6) dissolved in DMSO. Data represents 400-pixel averages from measurements with a 20 μs pixel dwell time. **d**. Spectral fidelity (left) and intensity-concentration relation (right) of serial diluted DMSO dissolved in DMSO-d6. Data represents 400-pixel averages from measurements with a 20 μs pixel dwell time. **e**. SRP lateral imaging of 100 nm PMMA beads in glycerol-d8 agar. Scale bar: 300 nm. **f**. Gaussian fitting of profile by the white dash line in e. **g**. SRP axial imaging of 100 nm PMMA beads in glycerol-d8 agar. Scale bar: 300 nm. **h**. Gaussian fitting of profile by the white dash line in g.

## 2.4 SRP preserves spectral fidelity under low NA light collection condition

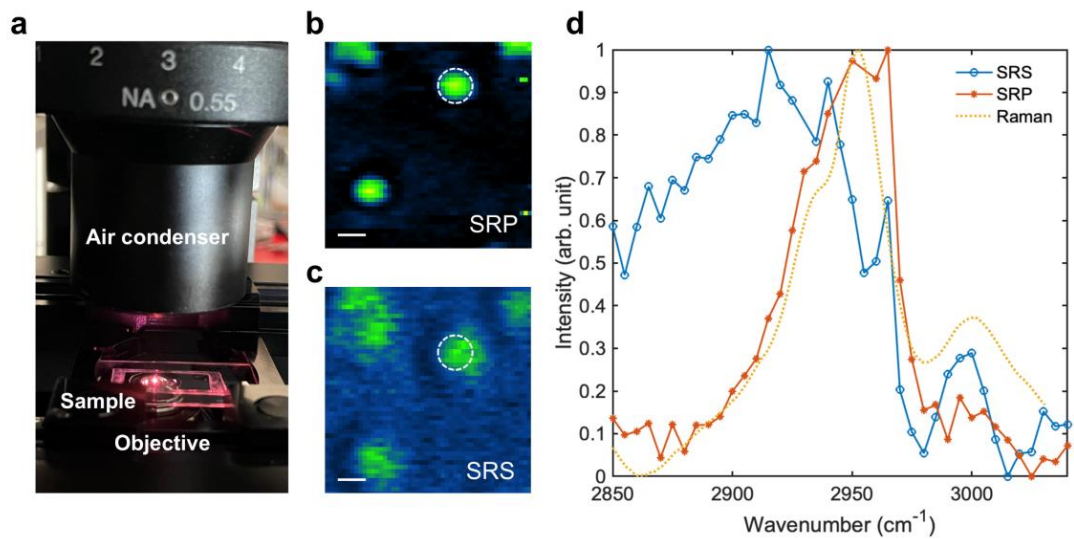

**Fig. 4** Spectral fidelity characterization of SRP and SRS under low NA light collection condition. **a**. Imaging condition of SRP or SRS (in stimulated Raman loss regime) using an air condenser. **b**. SRP imaging of 500 nm PMMA beads dried on the coverslip. Scale bar: 500 nm. **c**. SRS with autobalanced detection imaging of 500 nm PMMA beads dried on the coverslip with the same FOV as in **a**. Scale bar: 500 nm. **d**. Spectrum of PMMA beads measured by SRP and SRS in a and b in white dashed circles with the comparison to the spontaneous Raman spectrum of PMMA.

To demonstrate the capability of fiber laser SRP for imaging under low light collection conditions, we conducted spectral fidelity characterization of SRP and SRS (in stimulated Raman loss regime) microscopy. Using an air condenser with a NA of 0.55, we imaged 500 nm polymethyl methacrylate (PMMA) beads dried on a coverslip (**Fig. 4a**). The SRP imaging clearly resolved the beads and maintained their profile for size and shape (**Fig. 4b**). In contrast, the autobalanced-detected SRS image (**Fig. 4c**) of the same beads within the same field of view exhibited blurred spatial information and artifacts arising from other non-energy-absorbing parametric processes, such as four-wave mixing. The spectral data obtained from SRP and SRS imaging of the PMMA beads were compared to the spontaneous Raman spectrum of PMMA. The results (**Fig. 4d**) show that only SRP preserves the spectral information well under low NA light collection conditions, closely matching the spontaneous Raman reference. Collectively, these data demonstrate that SRP can maintain spectral fidelity and imaging quality when operating with low NA, large working distance light collection optics, making it suitable for various applications where thin or contact sample is not feasible.

## 2.5 SRP imaging cellular composition and dynamics in aqueous environment

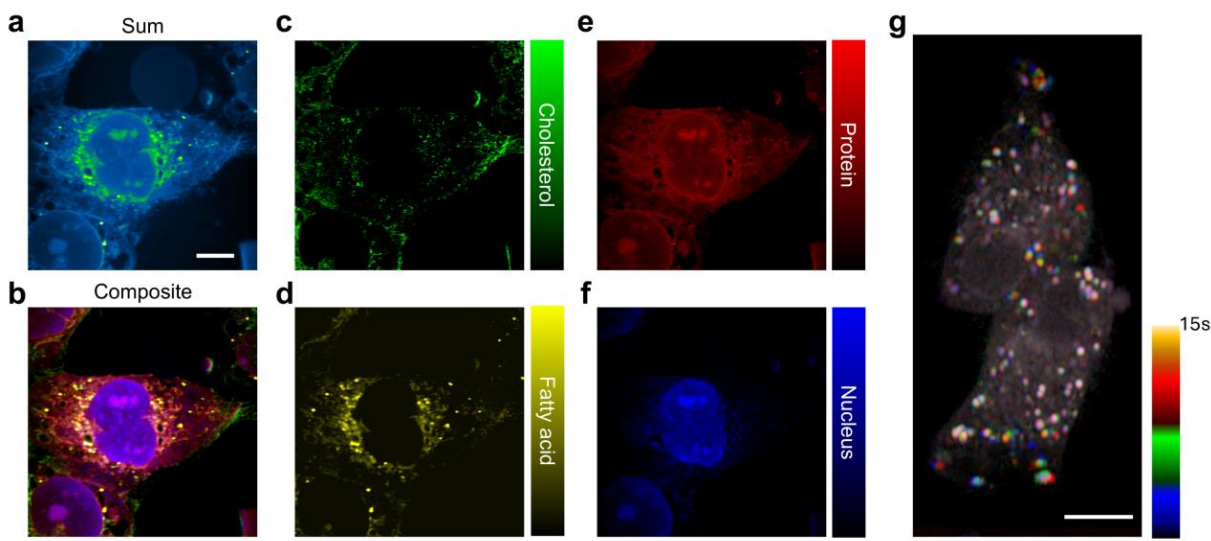

**Fig. 5** SRP imaging of cellular composition and dynamics in aqueous environment. **a-f**. Fixed cell imaging in PBS immersion condition with LASSO provided chemical maps. Scale bar: 8 μm. **g**. Live cell SRP imaging with color coded lipid movement at 2930 cm$^{-1}$, plot with logarithmic scale to feature weak features on membrane. Scale bar: 8 μm.

To show biological applications of fiber laser SRP, we applied our system to image intracellular cellular composition and dynamics of live cells in an aqueous environment, as demonstrated in **Fig. 5**. Samples were prepared in glass-bottom dishes with an open top, using phosphate-buffered saline (PBS) as the immersion medium to maintain cell shape and viability (see Materials and Methods). In SRS, cell imaging is often performed using an upright microscope equipped with a water immersion objective (**Fig. 1b**). The SRS setup typically requires the objective to be dipped into the sample to achieve the necessary working distance, which can lead to sample contamination. Additionally, the objective must be elevated to change samples, which disrupts the optical path and necessitates readjustments with each sample change. In contrast, our fiber laser SRP system is designed with an inverted microscope configuration (**Fig. 1b**), providing a 28 mm air gap between the sample and the light collection optics. This design allows for contact-free imaging, minimizing the risk of contamination and simplifying the process of changing samples without the need to adjust the optics.

Using hyperspectral SRP in the C-H region (2790 cm$^{-1}$ to 3030 cm$^{-1}$), we mapped key biochemical components—including cholesterol, proteins, lipids, and nuclear-specific regions—in fixed T24 bladder cancer cells. The chemical maps generated through least absolute shrinkage and selection operator (LASSO) analysis exhibited high spatial and chemical contrast [39], offering a detailed view of subcellular organization (**Fig. 5a-f**, **Fig. S5,** see Materials and Methods). Moreover, the fiber laser SRP system proved its robustness by enabling real-time imaging of live cell dynamics, specifically capturing lipid movement [40]. In live cell imaging, we achieved a speed of 1 frame per second with a 20 µs pixel dwell time, imaging a 200 by 200 pixels area. This setup allowed us to clearly visualize the live cells subcellular components, as well as the dynamic trajectories of lipids (**Fig. 5g**). The ability to perform live cell imaging while maintaining cell viability underscores the advantages of SRP over traditional methods, particularly in preserving cellular integrity and monitoring dynamic processes in vitro.

## 2.6 Highly sensitive SRP imaging of cell metabolism

Understanding metabolism at single cell level is crucial to biological research, and vibrational imaging has become a key technique for this purpose [1,41]. Fiber laser SRP offers a highly sensitive tool for studying cellular metabolism. We demonstrate that immersing biological samples in a biocompatible medium with low heat capacity and high thermo-optic coefficients, such as glycerol and urea, can significantly enhance SRP contrast. This enhancement allows for sensitive imaging of metabolites, particularly in the silent window where carbon-deuterium bond (C-D) stretching vibration resides.

**Fig. 6a-h** illustrates the application of SRP imaging to investigate lipid metabolism in oleic acid-d$_{34}$ (OA-d$_{34}$)-treated T24 bladder cancer cells (see Materials and Methods). The SRP images (**Fig. 6a-d**) of OA-d$_{34}$ treated T24 cells immersed in glycerol reveal subtle features in the cytoplasm and the membrane when captured with both parallel and orthogonal pump and probe beam polarizations at 2105 cm$^{-1}$ and 2205 cm$^{-1}$. The logarithmic scale plots highlight the contrast difference across different polarization conditions. The spectra of lipid droplet in white dashed

circle in **Fig. 6a-d** under these conditions is presented in **Fig. 6e**, showing the polarization dependence of the symmetric and asymmetric stretching bands, providing insights into the molecular orientation and environment within the cell [42].

We further explored the cell metabolic activity by imaging newly synthesized C-D bonds in the treated cells, using urea as the immersion medium, to monitor C-D and C-H in the same cell [33]. The silent region at 2105 cm$^{-1}$ (**Fig. 6f**) and the C-H region at 2930 cm$^{-1}$, corresponding to proteins and lipids (**Fig. 6g**), were used to reveal the distribution of these biomolecules. The overlay of these images (**Fig. 6h**) confirms the colocalization of newly synthesized lipids, demonstrating the capability of SRP to capture metabolic processes in both C-D and C-H regions simultaneously.

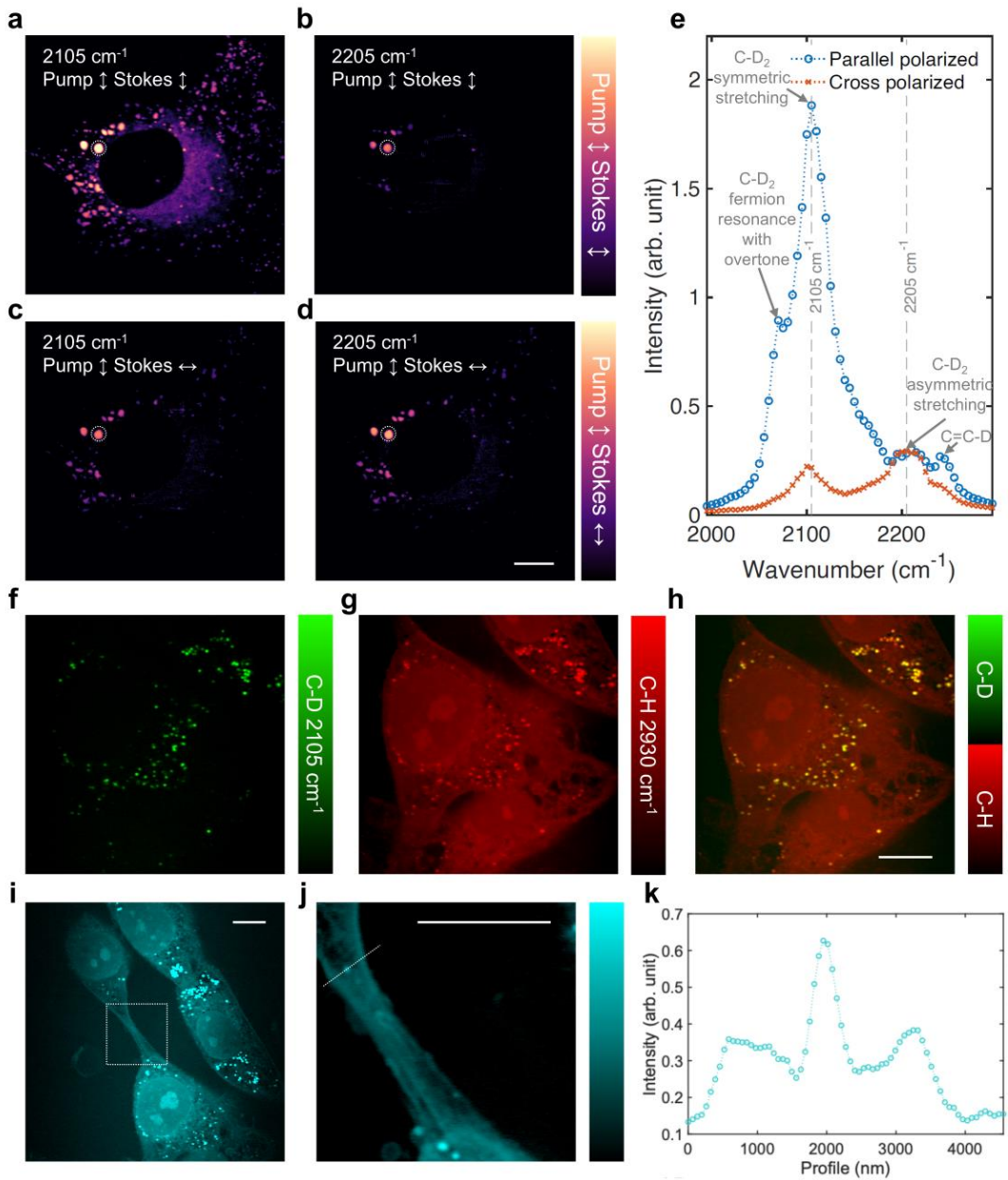

**Fig. 6** Silent window metabolism imaging by SRP in a photothermal enhancing medium. **a-d**. Oleic acid-$d_{34}$ (OA-$d_{34}$) treated T24 bladder cancer cells with parallel (a,b) or orthogonal (c,d) pump and probe beams excite at 2105 cm$^{-1}$ (a,c) or 2205 cm$^{-1}$ (b,d). Plot with logarithmic scale to feature weak features on the cytoplasm and membrane. The contrast is the same under same polarization condition. Scale bar: 10 μm. **e**. The spectra of the lipid droplet in the dashed circle in a-d inside OA-$d_{34}$ treated T24 cell under different polarization condition. **f**. Silent region imaging of the newly synthesized lipid droplets of the OA-$d_{34}$ treated T24 bladder cancer cell at 2105 cm$^{-1}$. Urea solution was used as the immersion medium. **g**. C-H region imaging of the same cell in f, mapping the protein and lipids distribution at 2930 cm$^{-1}$. **h**. Composite of f and g, show colocalization of the newly synthesized biomolecules. Scale bar: 10 μm. **i**. C-H region imaging at 2930 cm$^{-1}$ of a connected T24 cell immersed in urea solution. Scale bar: 10 μm. **j**. Zoom in of the box in **i**, showing detailed membrane connection and lipid allocation in the cell protrusion. Scale bar: 10 µm. **k**. Profile intensity of the line in **j**.

The advantages of SRP in thermal-enhancing media are further highlighted by the detailed imaging of two connected T24 cells (**Fig. 6i**). Immersion in urea solution allowed us to visualize intricate membrane connections and lipid distributions within cell protrusions (**Fig. 6j**). The intensity profile along a selected line (**Fig. 6k**) quantitatively confirms the enhanced signal achieved with urea, reinforcing its utility in SRP imaging.

Together, these findings highlight the sensitivity of SRP imaging in detecting subtle spectral differences, such as those observed in OA-d34-treated cells under different polarization excitations, while also demonstrating the capability for simultaneous C-D and C-H imaging. The use of the immersion medium further enhances sensitivity, enabling the detailed observation of fine cellular features, particularly in membranes and lipid droplets.

## 2.7 Volumetric SRP histology with tissue cleared rat brain

Vibrational chemical microscopy plays a crucial role in tissue imaging by enabling the label-free visualization of molecular structures, allowing researchers to capture intricate chemical compositions of biological tissues without the need for external dyes or markers [43,44]. This technique is especially valuable for revealing detailed spatial distributions of biomolecules, such as lipids and proteins, in both healthy and diseased tissues, offering profound insights into cellular processes and disease pathology at a subcellular level [5]. To demonstrate the SRP system's potential to achieve high-resolution, three-dimensional histological mapping of complex tissue structures, we implemented volumetric SRP imaging of urea-cleared rat brain slices. This approach leverages SRP's strong spatial and axial sectioning abilities. Moreover, the use of a tissue-clearing agent urea [33], not only improves the imaging depth but also increases signal sensitivity due to its favorable thermal properties compared to water. The developed fiber laser SRP system employs a sequential scanning strategy, beginning with spatial x-y scanning by galvo mirrors, followed by wavenumber-specific imaging, FOV adjustments for large area mapping, and finally depth scanning in the z-direction, as outlined in **Fig. 7a**.

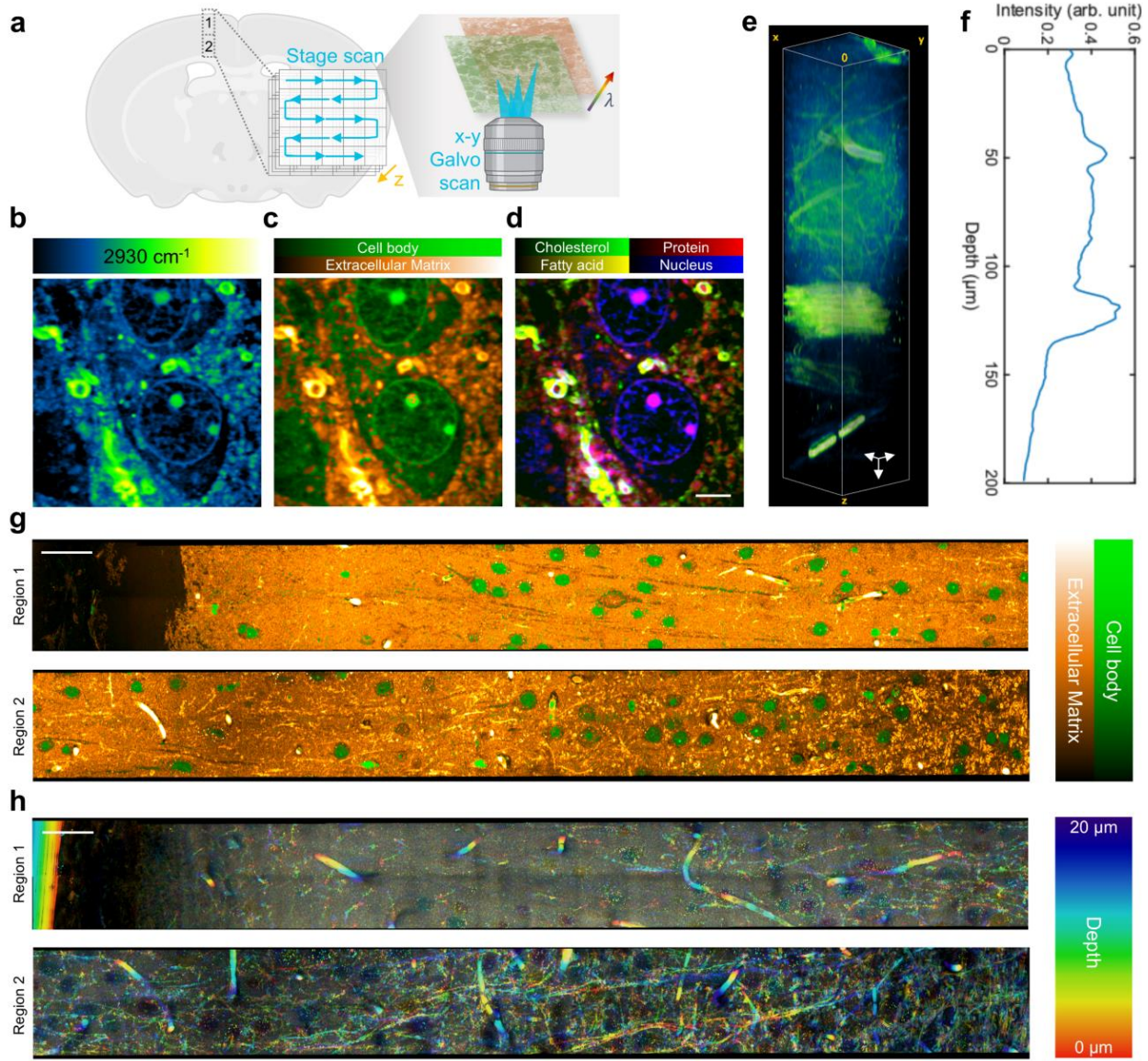

**Fig. 7** Volumetric SRP imaging of rat brain slice with tissue clearance. **a**. Illustration of volumetric SRP imaging sequence: spatial x-y (galvo scan)→ λ scan → large FOV (stage scan) → z scan. **b**. Single-color SRP image of a urea-cleared rat brain sample at 2930 cm$^{-1}$. **c**. Two-color SRP histology at the same FOV as in b. **d**. Biomolecule decomposition by hyperspectral SRP in C-H region. Scale bar: 5 μm. **e**. Depth-resolved SRP imaging of urea-cleared rat brain at 2930 cm$^{-1}$. Depth: 200 μm. Scale bar: 10 μm. **f**. Average intensity of each frame in the depth-resolved image in e. **g**. Two-color SRP histology of one layer in cerebral cortex. Region 1: V2MM outer region, closer to scalp. Region 2: V2MM inner region, close to corpus callosum. Locations are illustrated in a. Scale bar: 50 μm. **h**. Single-color depth-resolved SRP image of V2MM at 2930 cm$^{-1}$ in the same location as in g. Depth scan range: 20 μm. Scale bar: 50 μm.

The fiber laser SRP system accommodates multiple scanning modes, including single-color, two-color, and hyperspectral imaging, enabled by the fast wavelength tuning fiber laser source

(see Materials and Methods) [35]. Each mode provides distinct insights into the urea cleared brain's microarchitecture. The single-color mode at 2930 cm$^{-1}$ (**Fig. 7b**) highlights the general structure of the brain tissue by targeting C-H bonds prevalent in biomolecules, offering an overview of the biological matrix with fine resolution to even visualize the axions and dendrites of neurons (**Fig. S6**). We can see the different distribution of proteins and lipids in axon and dendrite, which is clear reflection of the fact that neurons have a highly differentiated structure with a preference for different molecule transportation. These regions are morphologically and functionally distinct, so the transport of substances must be precisely directed to specific locations to ensure proper nerve cell function. The two-color mode (2850 cm$^{-1}$ and 2930 cm$^{-1}$) facilitates the differentiation of cell bodies from the extracellular matrix by analyzing the ratio of these two signals, thus generating a detailed histological map of different brain regions (**Fig. 7c**). The hyperspectral mode, which fully samples the C-H stretching region, enables comprehensive biomolecular decomposition by referencing standard spectra of key components including cholesterol, proteins, fatty acids, and nucleic acids, thereby enhancing the specificity of cellular and subcellular mapping (**Fig. 7d** and **Fig. S7**). Furthermore, the fiber laser enabled fast scanning of the whole Raman spectral region (700-3200 cm$^{-1}$), including the fingerprint region (**Fig. S8**), which could provide more insight into subtype lipids and other biomolecules [45,46]. This level of detail is crucial in neuroscience, where distinguishing between different cell types, such as neurons and glial cells, offer a way for understanding functional brain organization and pathology [5].

The 300 μm thick rat brain tissue was cleared using an 8 M urea solution, following a protocol established by Mian Wei et al. [33] This approach extended the SRP imaging depth to over 200 μm (**Fig. 7e,f, Video S1**), constrained mainly by the working distance of the imaging objective. Urea, with its lower heat capacity relative to water, also contributes to the enhanced SRP signal strength, thereby improving overall imaging sensitivity. In **Fig. 7g-h**, the system's histological capabilities are exemplified through imaging the V2MM region of the rat brain, with distinct views of the outer region near the scalp (Region 1) and the inner region adjacent to the corpus callosum (Region 2). The two-color SRP histology clearly delineates cell bodies from the extracellular matrix (**Fig. 7g**), providing crucial information on neuron density across different brain layers (**Fig. S9**). Other brain functional regions are examined and showed distinct cell distributions (**Fig. S10**). Moreover, with depth-resolved information (**Fig. 7h, Video S2, Video S3**), a comprehensive view of the brain map could be visualized without the need of slicing the brain layer by layer. In particular, the myelin sheath and Schwann cell could be clearly visualized as it has high amount of lipids. To summarize, volumetric SRP provides a fast way visualizing the complex neural network inside the brain without the need for physical sectioning. Our results demonstrate that with tissue clearance, the fiber laser SRP system successfully extends the imaging depth, allowing clear visualization of features across different brain regions in a three-dimensional context. This approach not only captures fine subcellular structures but also offers a non-invasive method to explore intricate brain architectures, offering the potential of understanding neural functions and pathologies with high speed.

**2.8 Fiber laser SRP tissue imaging with low NA, long working distance objective**

Although the incorporation of low NA condenser has relaxed the sample format requirements on the detection side, the high NA objective on the illumination side still only allows shallow working distance. A low NA, long working distance objective in fiber laser SRP could provide sufficient space on both sides of the sample, making it more flexible for the format of samples. To demonstrate the compatibility of fiber laser SRP with low NA, objectives, we performed SRP imaging with a 0.45 NA, 6.6 mm working distance objective (Olympus LUCPLFLN20X). It is interesting to note that, with reduced illumination NA, the focal spot size of SRS excitation increases drastically, which leads to a reduced cooling rate. Therefore, a lower modulation frequency (60 kHz) is chosen to allow sufficient cooling and improve the signal. As shown in **Fig. S11a-b**, with 3 μm PMMA particle as a test bed, low NA fiber laser SRP could resolve the particles very well with the Raman spectrum at C-H region well captured. We then demonstrated low NA fiber laser SRP on mouse brain slice, as shown in **Fig. S11c-e**. The contrast is majorly from myelin sheath in the brain, where the protein rich (e.g. ROI1) and lipid rich (e.g. ROI2) regions could be well-differentiated from their spectrum. Notably, the very bright spots in the FOVs are arising from the linear absorption photothermal signals of residual red blood cells in brain blood vessels. Collectively, fiber laser SRP is compatible with low NA, long working distance objectives, allowing more flexibility sample formats.

## 3 Discussion

We have reported a fiber laser SRP microscope with high sensitivity, ease of operation, and broad compatibility with various biological samples. Fiber laser SRP offers 1~2 orders SNR improvement over SRS using the same fiber laser. Our results highlight the fiber laser SRP system's ability to perform metabolic analysis of cells in aqueous environment. Moreover, synergistic integration with tissue clearance agent as a thermal enhancing medium allows volumetric SRP imaging with enhanced depth and resolution, achieving three-dimensional histological mapping in cleared rat brain.

The fiber laser SRP system addresses a few challenges associated with SRS microscopy, including the laser intensity noise and the need for high NA light collection. Unlike SRS counterpart, which often suffers from noise in the excitation laser, fiber laser SRP leverages a third probe beam for thermal lensing detection, making the measurement less sensitive to noise in the pump and Stokes beams. This noise resilience simplifies the detection, eliminating the need for complex noise cancellation strategies such as autobalanced detection. Consequently, this fiber laser SRP system achieves a superior signal-to-noise ratio and sensitivity with a fiber laser. This technology offers several key strengths. First, the use of a fiber laser source makes the fiber laser SRP system compact, robust, and portable, suitable for both laboratory and clinical environments. Second, the rapid wavelength tuning capability, covering the Raman spectral range from 700 to 3200 $cm^{-1}$ within milliseconds, facilitates comprehensive chemical mapping, including functional studies and phenotypic linkage identification through fingerprint region imaging. Third, the inverted microscope configuration with low NA and long working distance optics allows for contact-free imaging, significantly reducing sample handling challenges and enabling easy integration with commercialized confocal microscopy setups. The system's compatibility with various advanced testing methods, such as flow cytometry, multi-well-plate-based assays, and cell ejection setups, enhances its versatility. These features collectively position the fiber laser SRP microscope as a platform for a wide range of applications, from single-cell analysis to tissue diagnostics.

There are spaces to further improve the system performance. First, due to the laser availability in lab, a narrow linewidth 765 nm CW laser with a long coherence length was used as the probe laser and introduced interference patterns when imaging samples with abrupt refractive index changes, such as at water-cell interfaces. This issue was mitigated by spatial oversampling and median filtering, while short coherence length laser could avoid this pattern from generation [30]. Second, the power limitations of the fiber laser used in the paper restrict the ability to employ lower duty cycles [30], which could enhance the SRP effect with the same average power on sample but higher energy deposition with higher peak power due to the nonlinearity of the SRS process. Adjusting the laser configuration, such as utilizing lower repetition rates or burst modes, could improve the sensitivity and overall performance of the SRP system [30]. Third, the heat caused from dissipative processes, such as multiphoton absorption, transient absorption, overtone absorption involved with the modulated beam will contribute to the background [47–49]. For the samples that suffer from other absorption processes, we only modulated pump or Stokes to flexibly avoid one of the beams contributed photothermal signal. For further reduction of the background, separating pump and Stokes pulse in time by delay modulation could eliminate the absorption from both beams. Post background removal will also be helpful, as other absorption features are mostly slow varying in the spectral domain [50].

In conclusion, the fiber laser SRP microscope represents a significant advancement in the field of vibrational imaging, offering high sensitivity, operational simplicity, and broad applicability. This system has the potential to facilitate novel applications including high-throughput screening, in vivo tissue diagnostics, and studies of cellular metabolism and dynamics. Further optimization of the laser source and detection setup will continue to enhance the system's performance, paving the way for its wider adoption in research labs and clinical practice.

## 4 Materials and methods

### 4.1 Fiber laser SRP microscope

A picosecond tunable fiber laser (Picus Duo, Refined Laser Systems) was used to generate synchronized pump and Stokes beams, with a tunable range covering vibrational frequencies from 700 $cm^{-1}$ to 3100 $cm^{-1}$ within 100 ms. It provided wide spectral coverage and fast wavenumber tuning speeds. It can be tuned to a wide range of wavenumbers in less than 100 ms, compared to over 10 s with solid-state lasers. The pump and Stokes beam powers were adjusted using a combination of half-wave plates and polarized beam splitters following the laser output. Two 1.5× Keplerian beam expanders were installed on the pump and Stokes beams, respectively, with acousto-optic modulators (1205c-1 with coating for pump or Stokes beam, Isomet Corporation) placed at the focal points of the expanders. A function generator controlled the modulation of the AOMs on the pump and/or Stokes beams. When both beams were modulated to lower the power on sample, modulations on both beams need to be in-phase to overlap the duty on period of pump and Stokes in time. The generator signal also served as a reference for the lock-in amplifier (HF2LI, Zurich Instruments) for subsequent signal demodulation. The probe beam (TLB6712-D, Spectral Physics) was expanded using a 2× Keplerian beam expander, with its second lens mounted on a manual delay stage. The probe beam was first combined with the pump beam via a polarized beam splitter, and then both were merged with the Stokes beam using a long-pass dichroic mirror (DMLP1000, Thorlabs). The combined beams were directed into a galvo system (GVSK2-US, Thorlabs), conjugated by a 3× expansion 4f system to the back aperture of the objective (UPlanApo 60XW, NA 1.2, Olympus) mounted on an inverted microscope (IX71, Olympus). A tunable NA air condenser (NA 0.55, Nikon) collected the transmission light through the sample. A 75 mm focusing lens then directed the collected light from the condenser onto a silicon photodiode (S3590-08) with a 64 V bias, equipped with mounted optical filters (FBH770-10, FESH1000, Thorlabs). The photocurrent was converted to voltage by a 50 ohms resistor (VT2, Thorlabs) and then filtered by 0.12 MHz high pass filter (ZFHP-0R12-s+, Mini-Circuits). The output alternative current voltage was pre-amplified (SA230-F5; NF Corporation) before sending to the lock-in amplifier for magnitude demodulation at the reference frequency provided by the function generator. The lock-in amplifier output was digitized by NI-DAQ card (PCIe-6363). LabVIEW (NI LabVIEW 2023 Q3, National Instruments) and MATLAB (R2023a, MathWorks) were used for the system control, synchronization and image real time display.

### 4.2 SRP thermal lensing effect simulations

The thermal lens effect in SRP microscopy was simulated by calculating the energy deposition based on the modulation depth of the SRS signal. The energy deposition data was used to model the localized heating within the sample. Assuming a 3D Gaussian Point Spread Function (PSF) to represent the SRS excitation, the resulting temperature distribution was derived, which characterizes the thermal lens effect. This 3D Gaussian-shaped thermal profile was generated using MATLAB by applying the energy deposition information to the Gaussian PSF model. The resulting thermal lens, due to different thermo-optic coefficient of the medium, was then exported as a spatial temperature distribution map.

Subsequently, the temperature-induced refractive index changes were mapped, and the data was imported into Lumerical FDTD (Finite-Difference Time-Domain) software (Ansys, USA). In Lumerical FDTD, the thermal lens profile was used to modify the refractive index distribution within the material, simulating the impact of the thermal lens on the optical wavefront. This enabled the calculation of the far-field light distribution, allowing for a detailed analysis of how the SRP-induced thermal effects influence light propagation and far-field patterns.

### 4.3 Beads sample preparation

Two different methods were employed to prepare the 100 nm bead samples used in this study. For dry bead samples, a 2 μL drop of 100 nm polymethyl methacrylate (PMMA) beads (MMA100, Degradex, Phosphorex, USA) was spread onto a coverslip and allowed to dry. Once dried, the sample was ready for SRP imaging on an inverted microscope.

For volumetric bead sample imaging in the C-H region, a 1% glycerol-d8-agar solution was prepared by mixing agar powder with glycerol-d8 and heating intermittently in a microwave until the agar was fully dissolved. The 100 nm PMMA beads were then added to the glycerol-d8 agar mixture once the temperature had lowered but before it solidified. A droplet of the glycerol-d8 agar and PMMA bead mixture was placed onto a coverslip, and another coverslip was sealed on top using double-sided tape as a spacer, creating a stable environment for volumetric SRP imaging.

### 4.4 Cancer cell lines, chemicals, and cell culture

T24 cells were obtained from the American Type Culture Collection (ATCC) and authenticated accordingly. The cells were cultured in high-glucose Dulbecco's modified Eagle's medium (DMEM, Gibco), supplemented with 10% fetal bovine serum (FBS, Gibco) and 100 U/ml penicillin/streptomycin (P/S). Cultures were maintained in a humidified incubator at 37°C with a 5% $CO_2$ atmosphere. For imaging experiments, cells were seeded into 35 mm glass-bottom dishes.

Oleic acid-$d_{34}$ (OA-$d_{34}$, CAS Number: 350671-54-4) was purchased from Sigma-Aldrich. To create a high-lipid environment model, OA-$d_{34}$ was dissolved in DMSO and then added to the culture medium at the desired concentrations. Cells were seeded for at least 24 hours before being exposed to fresh medium containing 100 μM OA-$d_{34}$, followed by a 24-hour incubation. For the control experiments, cells were treated with an equivalent volume of fresh medium without OA-$d_{34}$ for 48-hour.

### 4.5 Procedure of biological sample imaging in different immersion medium

For imaging fixed cells, the cells were first washed with PBS (1×, pH 7.4, Thermo Fisher Scientific) and then fixed with 10% neutral buffered formalin, followed by three washes with PBS. Unless otherwise specified, PBS was used as the imaging medium. For imaging in optimized photothermal media, PBS was removed from the culture dish and replaced with the appropriate immersion medium before imaging.

## 4.6 Preparation of rat brain slice with tissue clearance

All animal procedures adhered to the ethical guidelines and were approved by the Institutional Animal Care and Use Committee of Boston University. The experiment utilized an adult Long-Evans rat, which was anesthetized with 5% isoflurane before being euthanized for brain tissue collection. Transcardial perfusion was performed first with 1× phosphate-buffered saline (PBS, pH 7.4, Thermo Fisher Scientific) to remove blood, followed by 10% neutral buffered formalin (Sigma-Aldrich) to initiate tissue fixation. The brain was then extracted and further immersed in 10% neutral buffered formalin for 24 hours to ensure complete fixation. After fixation, the brain tissue was transferred to 1× PBS solution and sectioned into 300 μm thick coronal slices using an Oscillating Tissue Slicer (OST-4500, Electron Microscopy Sciences), making them suitable for imaging. The brain slices were subsequently immersed in a clearing solution of 8 M urea (U5378, Sigma-Aldrich) supplemented with 0.2% Triton X-100 (X100, Sigma-Aldrich) for more than 3 d at room temperature, following the protocol described by Mian Wei et al. [33], prior to imaging.

## 4.7 LASSO-Based Spectral Unmixing for SRP Hyperspectral Imaging

To analyze the SRP hyperspectral imaging data of fixed T24 bladder cancer cells, we employed a LASSO algorithm for spectral unmixing. This approach was chosen for its ability to handle the complex overlapping spectra typical of biological samples, allowing for the decomposition of SRP spectra into distinct contributions from specific molecular components. The reference spectra used in the LASSO analysis included key biochemical components relevant to cancer cell metabolism: cholesterol, bovine serum albumin (BSA) for protein, triacylglycerol (TAG) for fatty acid, and nuclear-specific signals. These reference spectra were acquired independently from pure samples, ensuring they accurately represent the molecular fingerprints within the SRP spectra.

The LASSO model was employed to solve the spectral unmixing problem by minimizing the residual sum of squares subject to an L1-norm constraint. This approach effectively suppresses non-relevant spectral contributions, enhancing the accuracy of the resulting concentration maps. Let $N_\lambda$ denote the wavenumber dimension of the SRP hyperspectral image and $K$ represent the number of pure components. The mathematical formulation of the LASSO problem is given by:

$$min_C \frac{1}{2}\|D - C \cdot S\|_2^2 + \beta\|C\|_1$$

where $D \in \mathbb{R}^{N_\lambda}$ represents the observed SRP spectrum at each pixel, $S \in \mathbb{R}^{K \times N_\lambda}$ is the matrix of reference spectra, $C \in \mathbb{R}^K$ is the concentration vector corresponding to each reference, and $\beta \in \mathbb{R}$ is the regularization parameter that controls the sparsity of the solution. The parameter $\beta$ was empirically optimized to ensure accurate spectral decomposition while suppressing non-specific signals and minimizing the impact of background noise. The LASSO analysis was applied on a pixel-by-pixel basis, yielding concentration maps for each molecular component. These maps provided spatially resolved information on the distribution of cholesterol, proteins (BSA), fatty acid (TAG), and nuclear components within the fixed T24 bladder cancer cells.

## Acknowledgements

### Availability of data and materials

The data generated and/or analyzed during the current study are available from the corresponding author upon reasonable request.

### Funding

This work is supported by NIH grants R35GM136223, R01EB032391, R01EB035429 to JXC.

### Author contributions

J.-X.C., X.G., Y.Z. conceived the idea. X.G. and H.N. designed and constructed the setup. X.G., D.S. and conducted the experiments. D.S., Y.L, and C.V.P.D. contributed resources. X.G. analyzed the data. X.G. and J.-X.C. wrote the manuscript. All authors read and discussed the results.

## Declarations

### Competing interests

The authors declare no competing interests.

# Supplementary Information for:

## Fiber laser based stimulated Raman photothermal microscopy with long working distance optics

Xiaowei Ge[a,†], Yifan Zhu[a,†], Dingcheng Sun[b], Hongli Ni[a], Yueming Li[a], Chinmayee V. Prabhu Dessai[b], Ji-Xin Cheng [a,b,c,d]*

[a]Department of Electrical & Computer Engineering, Boston University, Boston, Massachusetts, USA.

[b]Department of Biomedical Engineering, Boston University, Boston, Massachusetts, USA.

[c]Department of Chemistry, Boston University, Boston, Massachusetts, USA.

[d]Photonics Center, Boston University, Boston, Massachusetts, USA.

[†]These authors have contributed equally to this work

*Corresponding authors: jxcheng@bu.edu

**This PDF file includes:**

Figs. S1 to S11

Videos S1 to S3

Tables S1

Supplementary Method

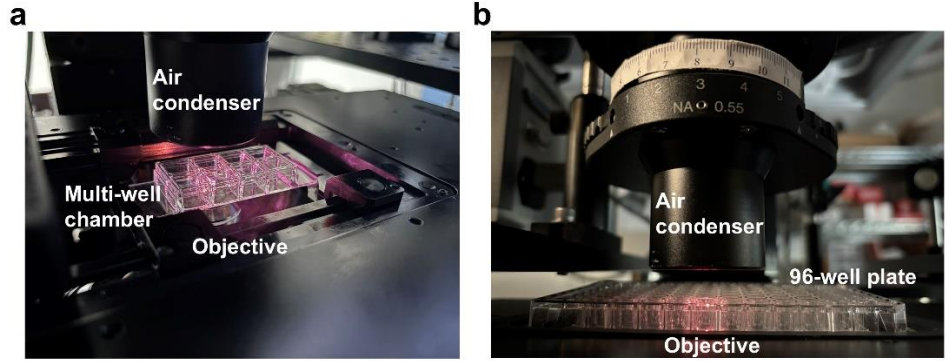

**Fig. S1** Experimental setup of SRP. **a**. Long working distance allows the test of multi-well chamber. **b**. Compatibility with 96-well plate.

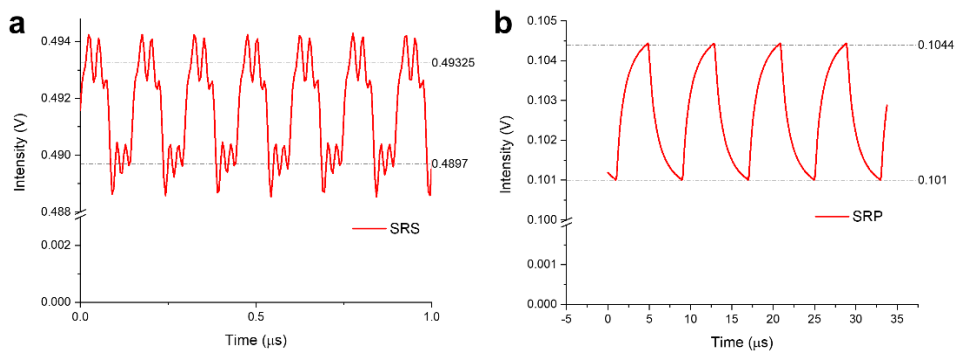

**Fig. S2** Modulation depth compare between SRS and SRP on DMSO. Pump power on sample: 28 mW, no modulation. Stokes power on sample: 90 mW, 50% duty cycle modulation. Probe power on sample: 23 mW. **a**. SRS modulation depth 0.72%. **b**. SRP modulation depth 3.3%.

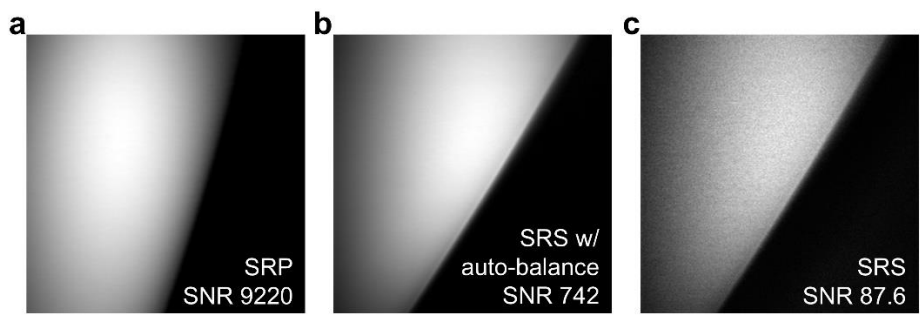

**Fig. S3** SNR comparison of SRP (**a**), SRS with auto-balance detection (**b**) and SRS without balance detection (**c**) on DMSO. Pump power on sample: 28 mW, no modulation. Stokes power on sample: 90 mW, 50% duty cycle modulation. Probe power on sample: 23 mW.

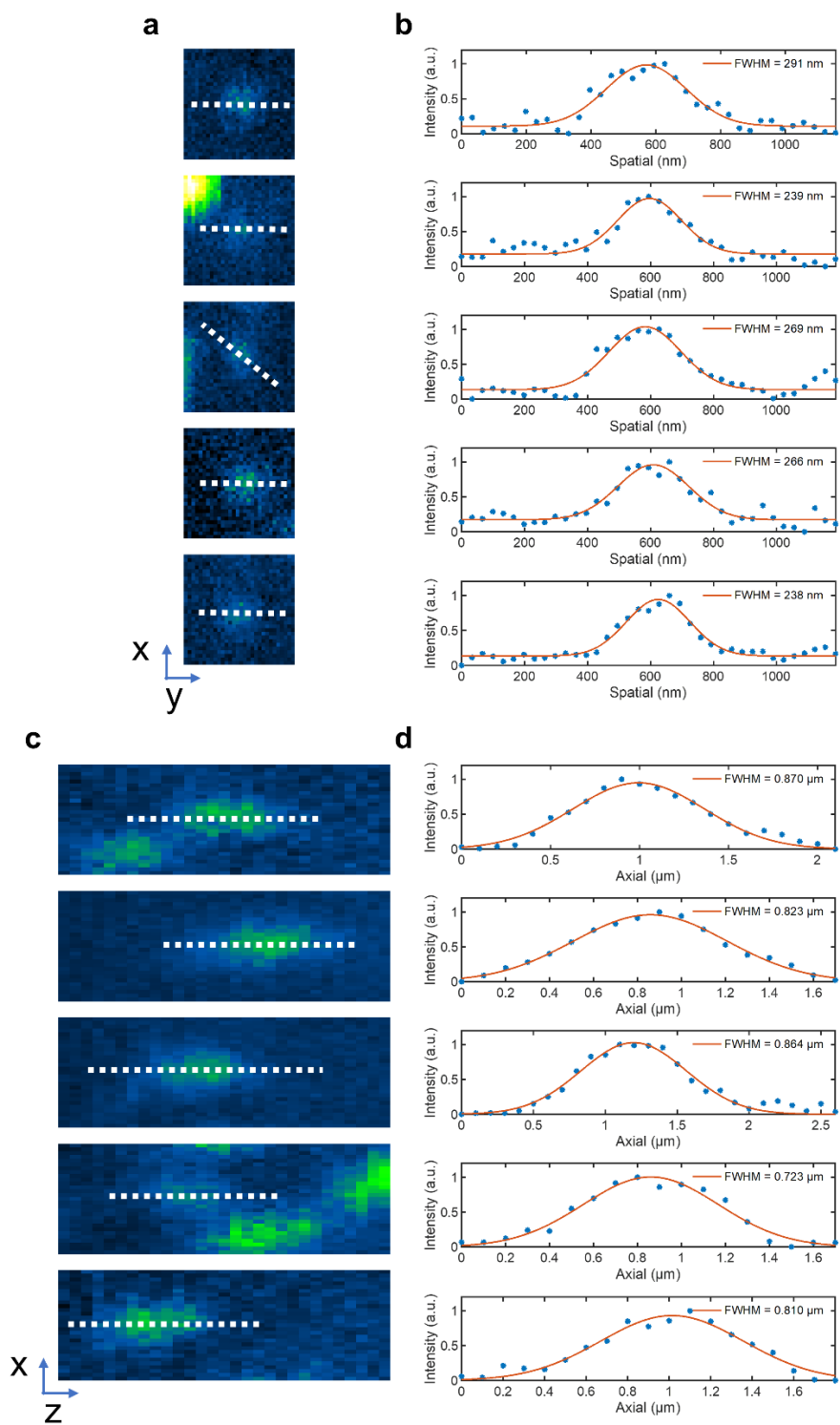

**Fig. S4** SRP spatial resolution characterization. **a**. SRP lateral imaging of 100 nm PMMA beads in glycerol-d8 agar. Scale bar: 300 nm. **b**. Gaussian fitting of profiles by the white dash lines in a. **c**. SRP axial imaging of 100 nm PMMA beads in glycerol-d8 agar. Scale bar: 300 nm. **d**. Gaussian fitting of profiles by the white dash lines in c.

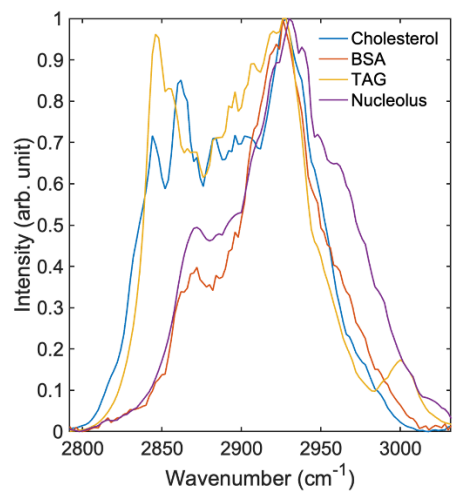

**Fig. S5** LASSO reference from standard sample SRP measurement. BSA: bovine serum albumin. TAG: triacylglycerol.

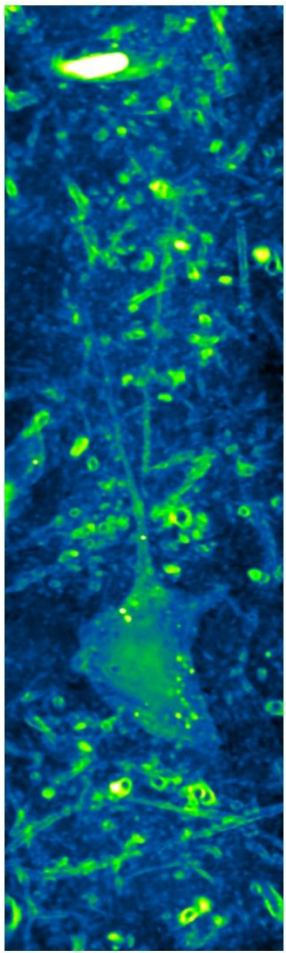

**Fig. S6** Single-color SRP image visualizing a single neuron with its axon and dentrite embedded in a urea-cleared rat brain at 2930 cm$^{-1}$. Scale bar: 10 μm.

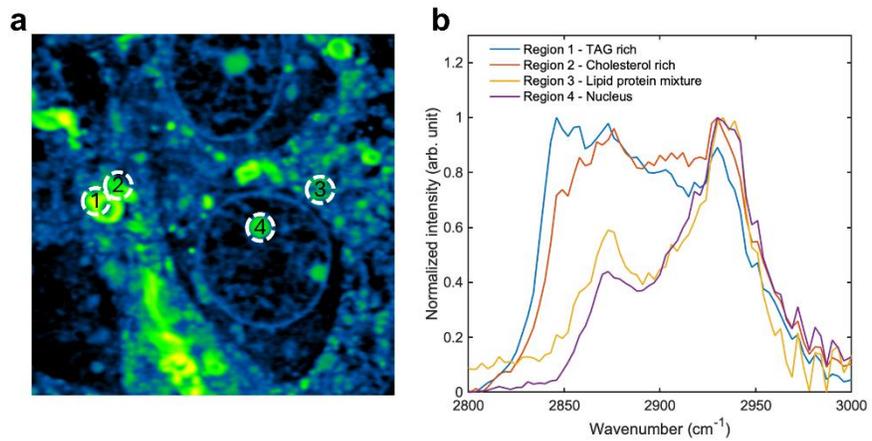

**Fig. S7** Hyperspectral SRP measurement of the tissue clearance rat brain sample. **a**. Single frame of hyperspectral SRP imaging at 2930 cm$^{-1}$. **b**. SRP spectrum of region of interests in a.

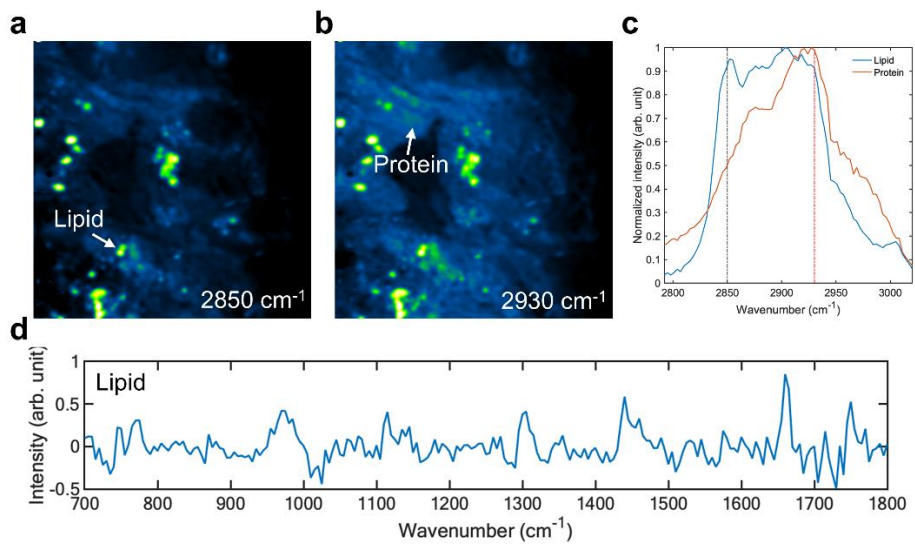

**Fig. S8** Ovarian tumor imaging in whole spectral region. **a.** SRP imaging of glycerol-d8 immersed tumor tissue at 2850 cm$^{-1}$. **b**. SRP imaging of the same FOV in a at 2930 cm$^{-1}$. **c**. C-H region SRP spectrum of lipid and protein pointed in a and b. Gray dashed line at 2850 cm$^{-1}$. Red dashed line at 2930 cm$^{-1}$. **d**. Fingerprint region SRP spectrum of lipid pointed in a.

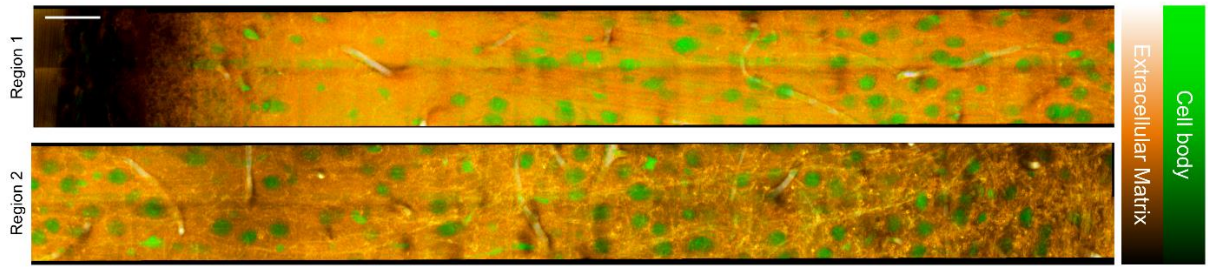

**Fig. S9** 20 µm depth projection of volumetric two-color SRP histology of cerebral cortex. Region 1: V2MM outer region, closer to scalp layer. Region 2: V2MM inner region, close to corpus callosum. Location illustrated in Fig. 7a. Scale bar: 50 µm.

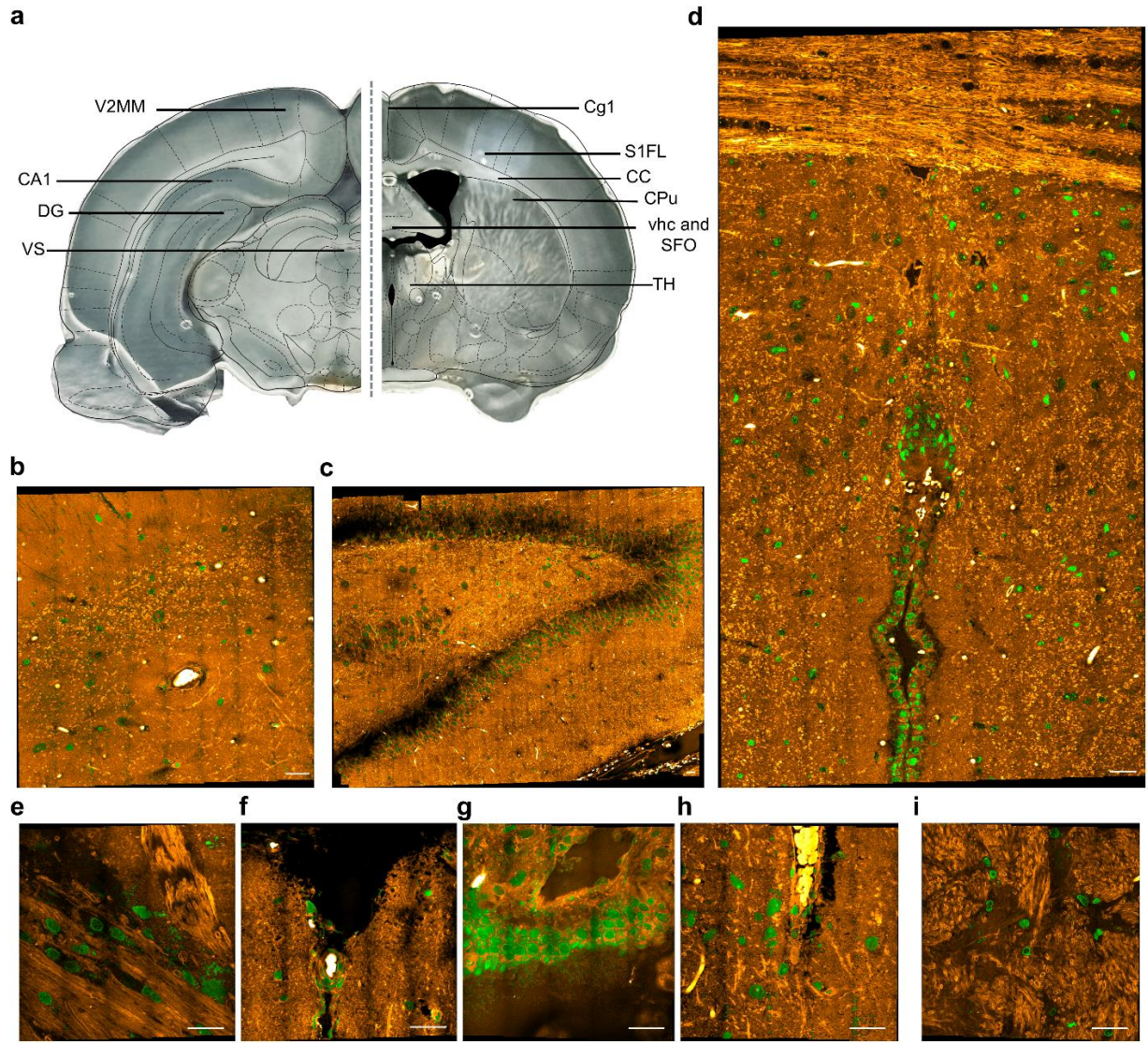

**Fig. S10** Two-color brain histology by fSRP. Scale bar: 30 μm. **a**. Image of the imaged rat brain slices. The left slice is close to the lambda and the right slice is close to the bregma. Two-color SRP histology of cornu ammonis 1 field in hippocampus (CA1) (**b**), dentate gyrus in hippocampus (DG) (**c**), ventricular system (VS) (**d**), primary somatosensory cortex, forelimb region in cerebral cortex (S1FL, top right) and corpus callosum (CC, bottom left) (**e**), cingulate cortex, area 1 in cerebral cortex (Cg1) (**f**), hippocampal commissure (vhc) and subfornical organ (SFO) (**g**), thalamus (TH) (**h**), caudate putamen in cerebral nuclei (CPU) (**i**).

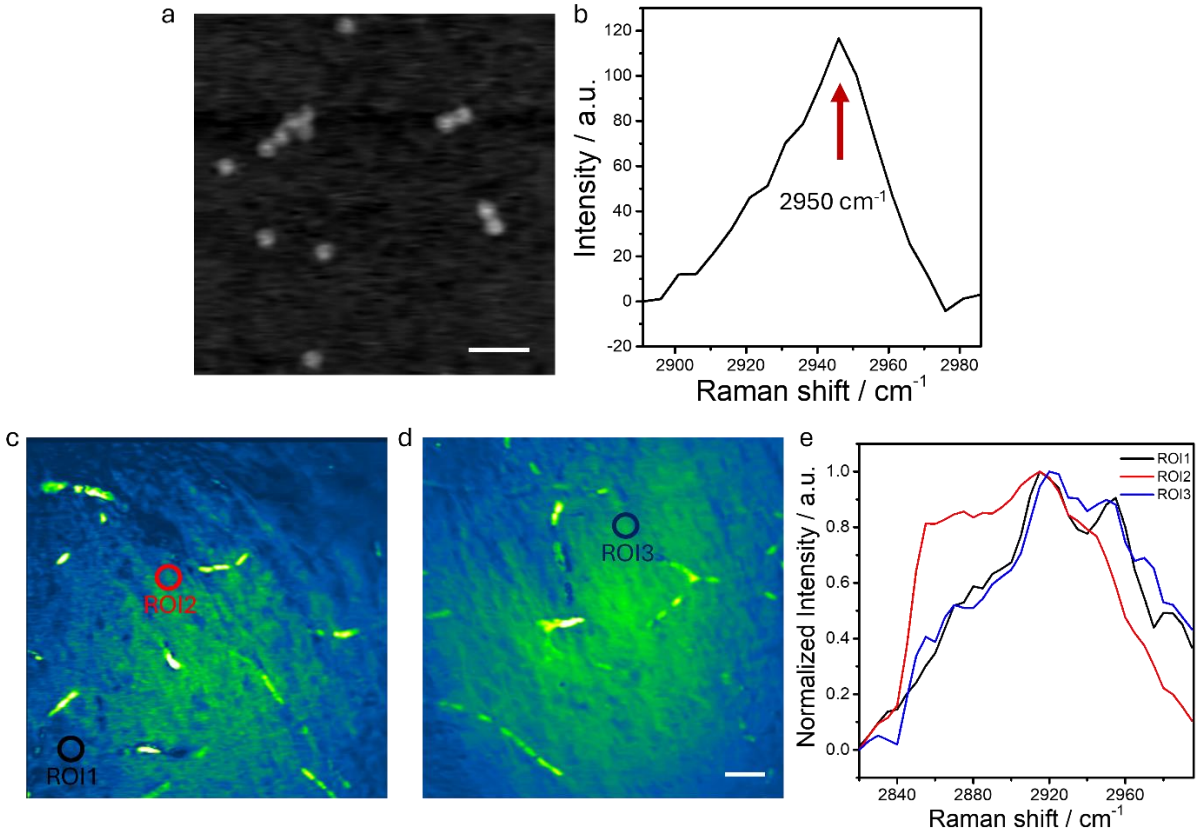

**Fig. S11** Hyperspectral imaging of 3 µm PMMA particles and mouse brain slice by low-NA fSRP. (**a-b**). low-NA fSRP image (**a**) and spectrum (**b**) of 3 µm PMMA particles. Scale bar: 10 µm. (**c-d**). low-NA fSRP images of mouse brain slice, at 2930 cm$^{-1}$. Scale bar: 20 µm. (**e**). typical SRP spectrum acquired at labeled FOVs.

**Video. S1** Live cell imaging in water condition with lipid movements as shown in Fig. 5g. Plot in logarithmic scale to feature weak features on membrane. Scale bar: 4 µm.

**Video. S2** Single-color depth-resolved SRP imaging of rat brain slice at 2930 cm$^{-1}$. Same FOV as Fig. 7e. Imaging depth: 200 µm. Scale bar: 5 µm.

**Video. S3** Two-color depth-resolved SRP histology of V2MM region (stitching of region 1 and 2 in Fig. 7) in rat cerebral cortex. Imaging depth: 20 µm. Scale bar: 50 µm.

| Thermal property | Unit | DMSO | glycerol | 8 M Urea | water |
|---|---|---|---|---|---|
| Heat Capacity | J/(kg·K) | 1966 | 2400 | 1420 | 4184 |
| Thermal conductivity | W/(m·K) | 0.200 | 0.283 | - | 0.598 |
| Thermo-optic coefficient dn/dT ($10^{-4}$) | $K^{-1}$ | -4.93 | -2.30 | - | -1.13 |
| Refractive index | - | 1.479 | 1.473 | 1.4 | 1.333 |
| Relative signal intensity | a.u. | 8.37 | 3.21 | | 1 |
| Viscosity | cP | 1.991 | - | 1-2 | 0.89 |

**Table S1** Thermal property of the immersion medium used in the study.

## Supplementary Method

**Preparation of rat brain slice with tissue clearance**

A fresh ovarian tumor was harvested from a euthanized NU/J mouse inoculated with OVCAR5-cisR cells (4-week-old female, homozygous for Foxn1nu, from the Jackson Laboratory). The tumor was immediately fixed in a 10% formalin solution. For cryopreservation, the tissue block was washed with 1× PBS solution (pH 7.4, Thermo Fisher Scientific) and incubated in a 15% sucrose solution for 12 hours. It was then immersed in a 30% sucrose solution overnight at room temperature until the tissue sank. The sample was embedded in optimal cutting temperature (OCT) compound (Fisher Healthcare) and stored at -80°C in a tissue mold prior to sectioning. Tissue slicing was performed using a Leica CM1950 cryostat at the Bio-Interface and Technologies Facility, Boston University. Sections of 10 µm thickness were placed on super frost glass slides and stored at -80°C until imaging. Before imaging, the slides were washed with 1× PBS to remove residual OCT and covered with fresh 1× PBS solution. The sample was then embedded in glycerol-d8 and covered with a glass coverslip for imaging shown in Fig. S8.